\documentclass[
    ,final            % use final for the camera ready runs
  ]
  {aipproc}

\layoutstyle{6x9}

\usepackage{natbib}

\newcommand{\msun}{\mbox{$M_{\odot}$}}

\newcommand{\zsun}{\mbox{$Z_{\odot}$}}

\newcommand{\Teff}{\mbox{$T_{\rm eff}$}}

\newcommand{\ratio}{\mbox{$v_{\infty}$/$v_{\rm esc}$}}
\newcommand{\mdot}{\mbox{$\dot{M}$}}

\begin{document}

\title{Mass loss predictions for hot stars}


\author{Jorick S. Vink}{
  address={Armagh Observatory, College Hill, Armagh, BT61 9DG, Northern Ireland (UK)}
}

%\author{<author2>}{
%  address={<common address for author2 and author3>}
%}

%\author{<author3>}{
%  address={<common address for author2 and author3>}
%  ,altaddress={<author1 address>} % additional visiting address
%}

\begin{abstract}
I present the results of radiation-driven mass-loss predictions for hot stars of all mass.
Mass loss is an important aspect for the evolution of massive stars, the rotational properties 
of the progenitors of gamma-ray bursts, and is essential in assessing whether the most massive stars 
explode as pair-instability supernovae, or avoid them due to mass loss. 
As a result, the rate of mass loss is critical for our understanding of the 
chemical enrichment of the Universe. Of particular interest is the question whether luminous blue variables are 
the {\it direct} progenitors of some supernovae. Although there is a growing body of evidence to suggest 
this, it remains as yet unexplained by state-of-the-art stellar evolution models. 
Finally, I discuss the relevance of mass loss for the appearance and rotational properties of 
hot Horizontal Branch stars in globular clusters and subdwarf B stars in the field.

\end{abstract}

\maketitle

%%%%%%%%%%%%%%%%%%%%%%%%%%%%%%%%%%%%%%%%%%%%
%% MAINMATTER
%%%%%%%%%%%%%%%%%%%%%%%%%%%%%%%%%%%%%%%%%%%%

\section{Introduction}

The occurrence of stellar winds is ubiquitous for stars of almost all mass and constitutes 
a key aspect of stellar physics. In terms of the mass loss integrated over a star's life, it is most extreme 
for the most luminous stars, where the winds are almost certainly driven radiatively. 
Therefore, I first describe the radiation-driven winds of massive stars, highlighting the implications for 
the least understood phases of their evolution, e.g. the luminous blue variable (LBV) phase.

The relevance of radiation-driven winds is however not limited to massive 
stars, but radiation-driven winds are thought to play a role for all types 
of hot stars, such as the post-AGB stars. 
Furthermore, certain types of main-sequence A and B stars (the magnetic Ap/Bp stars) show intriguing 
abundance patterns, which is the result of diffusion in the presence of a weak stellar wind. 
Similar diffusion and wind effects are likely operating in subdwarf B (sdB) stars and 
blue horizontal branch (BHB) stars.

But there is more: winds remove not only mass, but also angular momentum.  
For low-mass stars, I discuss the role of stellar winds for the rotational properties of
BHB stars in globular clusters.
For massive stars, angular momentum loss is particularly relevant with respect to the long-duration 
gamma-ray bursts (GRB) phenomenon, as the popular collapsar model requires the GRB progenitor to 
be rotating rapidly. 

Finally, I describe the role of metallicity $Z$, a key parameter in the physics of stars and galaxies, 
largely via the metallicity dependence of radiation-driven winds during the main sequence and final 
phases of stellar evolution.

\section{Predictions for massive OB stars}

The evolution of a massive star, with $M$ $>$ 30 \msun, is largely dominated 
by mass loss. The theory of radiation-driven winds was developed 
in the early 1970s by Lucy \& Solomon (1970) and Castor, Abbott \& Klein (1975, CAK).
Metal ions, such as Fe, are efficient scatterers of photons at specific line frequencies, and 
when the resulting radiative acceleration becomes larger than the inward pointing 
gravitational acceleration, an outflow results. For the dense winds of OB 
supergiants, Coulomb coupling is highly efficient, and the flow may be approximated
as a single fluid (although this assumption may break down for the weaker winds of Horizontal Branch stars).

\begin{figure}
\includegraphics[height=.3\textheight]{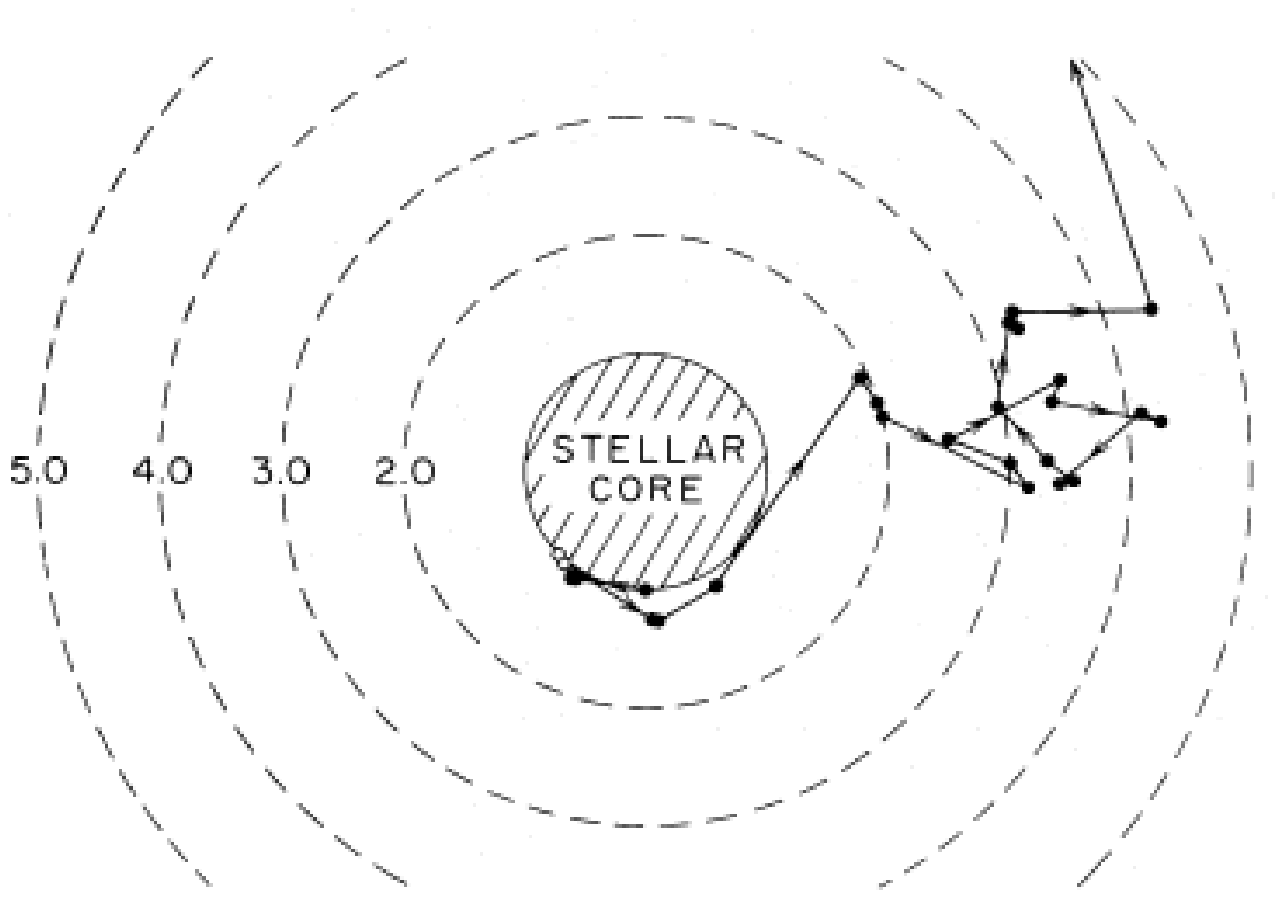}
\caption{The path of a typical Monte Carlo photon that undergoes multiple interactions
with matter on its way through the stellar wind, thereby depositing both energy and momentum.
Taken from Abbott \& Lucy (1985).}
\label{f_abbott}
\end{figure}

The early radiation-driven wind models and subsequent non-LTE improvements by e.g. Kudritzki \& 
Puls (2000) have been very successful in reproducing many wind properties.  
For dense winds however, these modified-CAK models do not suffice, as 
multiple scatterings are not included (e.g. Abbott \& Lucy 1985). 
Vink et al. (1999,2000) predicted the mass-loss rates of OB supergiants as a function
of stellar parameters including the effect of multiple scatterings (on line and continuum
opacity) using a Monte Carlo approach (see Fig.~\ref{f_abbott}). 
The mass-loss rates were found to scale as:

\begin{equation}
\dot{M}~\propto~L^{2.2}~M^{-1.3}~\Teff^1~(\ratio)^{-1.3}
\label{eq_formula}
\end{equation}
This shows that the mass-loss rate scales strongly
with luminosity ($L^{2.2}$).
The reason is that the more luminous stars have denser winds and the MC predictions  
deliver an increasingly larger mass-loss rate than modified-CAK predictions.
The results of Eq.~(\ref{eq_formula}) are being widely used in models 
of massive star evolution \footnote{they can be obtained from \url{http://www.arm.ac.uk/~jsv/}}.

\begin{figure}
  \includegraphics[height=.3\textheight]{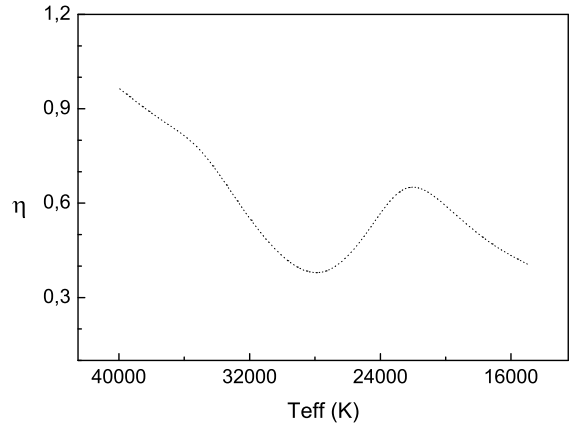}
  \caption{Predictions of the wind efficiency $\eta$ $=$ $\frac{\dot{M} v_{\infty}}{L/c}$ as a function of effective temperature.
Note the presence of a local maximum around 25~000 K, the bi-stability jump.}
  \label{vink99}
\end{figure}

Figure~\ref{vink99} shows the predictions as a function of effective temperature, expressed as 
the efficiency of the momentum transfer, $\frac{\dot{M} v_{\infty}}{L/c}$.
The plot shows a declining wind efficiency with effective temperature. 
At the higher O star temperatures, of $\sim$ 40~000 K, the flux and the opacity show a good ``match'' 
and the momentum transfer efficiency is high. When the temperature drops, the flux moves gradually 
towards lower wavelengths, resulting in a growing mismatch between the flux and the line opacity.

Around 25~000 K, a jump in the mass-loss rate is noted due to an increased Fe 
opacity. This ``bi-stability jump'' (Pauldrach \& Puls 1990) may recently have been confirmed in radio data that appear to confirm the 
presence of a local maximum (Benaglia et al. 2007), but the predictions below the jump temperature are much larger than 
predicted (see the discussion in Vink et al. 2000, Trundle \& Lennon 2005, Crowther et al. 2006). 

The bi-stability jump occurs where winds change from 
a low $\dot{M}$, fast wind, to a high $\dot{M}$, slow wind, due to a change in the Fe ionisation that 
drives the wind.  It may be an important ingredient for stellar evolution calculations, when stars evolve off 
the main sequence towards the red part of the Hertzsprung-Russell diagram (HRD). This is 
not only relevant for their {\it mass} loss, but also for the loss of {\it angular momentum}. 
It may also play a role for LBV mass loss.

\begin{figure}
  \includegraphics[height=.3\textheight]{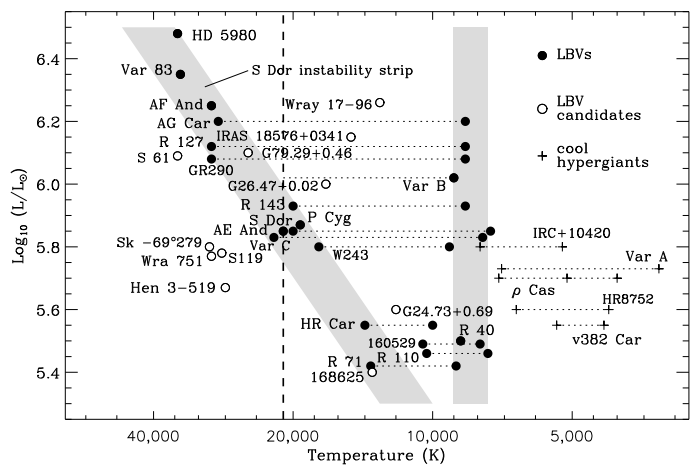}
  \caption{The position of luminous blue variables in the HRD. The shaded areas represent the objects during visual 
minimum (the tilted area on the left) and maximum (the straight vertical area on the right 
at $\sim$ 8000 K). The dashed vertical line at 21000 K is the position of the 
bi-stability jump. The figure is adapted from Smith et al. (2004).}
  \label{smith}
\end{figure}

\section{The winds of Luminous Blue Variables}

Figure~\ref{smith} shows the HRD position of the confirmed galactic and Magellanic Cloud 
LBVs. The objects change their effective temperatures (and stellar radii)
on a variety of timescales with various amplitudes (e.g. Humphreys \& Davidson 1994).
The so-called {\it micro} variations are intriguing but similar to other supergiants 
in this part of the HRD.
At the other extreme are the Eta Car-type huge eruptions, also referred to as SN-impostors when they occur in 
external galaxies (e.g. Van Dyk et al. 2000). We note that only two giant outbursts have ever been recorded 
in the Galaxy (the eruptions of Eta Car in the 19th and P Cyg in the 17th century). 
Most typifying for LBVs as a class are the S~Dor variations. 
For this variability, the amplitudes vary on timescales of years to decades (e.g. van Genderen 2001).

\begin{figure}
  \includegraphics[height=.3\textheight]{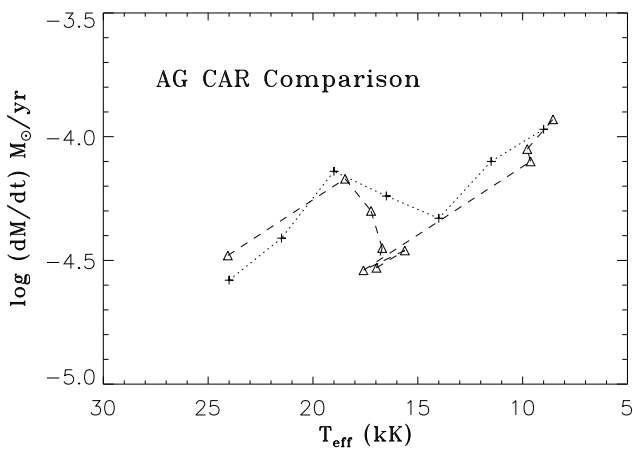}
  \caption{The dashed line represents variable mass-loss predictions for LBVs vs. effective temperature. 
Also shown (dotted line) are the empirical mass-loss rates for AG~Car analysed by Stahl et al. (2001).}
  \label{ag}
\end{figure}

When LBVs such as AG Car -- one of the prototype S~Dor variables -- change their 
radii on their S~Dor timescales, they show large mass loss variations (Stahl et al. 2001). 
Such variable wind behaviour has qualitatively been reproduced by radiation-driven wind 
models of Vink \& de Koter (2002) as depicted in Fig.~\ref{ag}. 

We anticipate that this type of wind behaviour may result in a circumstellar medium consisting of concentric shells with varying 
densities. This may have ramifications for the end-points of massive stars. 
Kotak \& Vink (2006) recently suggested that the quasi-periodic modulations seen in the radio lightcurves of some supernovae (SNe), 
such as 2001ig (Ryder et al. 2004) and 2003bg (Soderberg et al. 2006) may indicate that LBVs could be the {\it direct}
progenitors of some SNe. At first this seems to contradict stellar evolution calculations, which do not predict
LBVs to explode, and such a scenario was until recently considered ``wildly speculative'' (Smith \& Owocki 2006). 
Nevertheless, the recent case of SN 2006jc that showed a giant eruption just 2 years before it exploded 
(Pastorello et al. 2007, Foley et al. 2007) may add confidence to our suggestion, although there are unresolved issues with 
the chemical abundance of 2006jc and its progenitor. 
There are other tantalising indications that LBVs may explode. Gal Yam et al. (2007) may have detected 
a very luminous progenitor of SN 2005gl, although it remains to be seen if this is a single object. Although the properties 
of the potential progenitor star are consistent with that of an LBV, a hypergiant cannot be classified as an LBV 
until it has shown S~Dor or Eta-Car-type variability. A final interesting hint that LBVs may explode comes from the striking similarities 
in LBV nebula morphologies and the circumstellar medium of SN 1987A (Humphreys \& Davidson 1994, Smith 2007).
%These range from the possible detection of a 
%very luminous progenitor of SN 2005gl (Gal Yam et al. 2007) to 
%similarities in LBV nebula morphologies and the circumstellar medium of SN 1987A (Smith 2007). 
%Furthermore, the most luminous supernova ever recorded, SN 2006gy (Smith et al. 2007, Ofek et al. 2007) may also have been an LBV.

As current state-of-the-art stellar evolution calculations do not predict LBVs to explode, this 
represents a major unresolved problem in the physics of massive stars.

\section{Stellar winds on the Horizontal Branch}

Moving our discussion to a group of lower mass evolved hot stars, we arrive at objects on the HB. HB stars are 
low-mass stars with a helium core of about half a solar mass and a small hydrogen layer on top. 
Vink \& Cassisi (2002) provided maximal (where the terminal wind velocity equals the escape velocity) 
mass-loss predictions for HB stars:

\begin{equation}
\dot{M}~\propto~L^{2.1}~M^{-1.1}~\Teff^1~Z^1
\label{eq_formula2}
\end{equation}
The predicted rates may account for the remaining mass-discrepancy in 
BHB stars (Vink \& Cassisi 2002, Moehler et al. 2003) whilst 
at the same time accounting for the observed H$\alpha$ profiles of luminous SdB stars (Vink 2004, Heber et al. 2003).
Stellar winds may also be required to explain the complex abundance patterns found 
in these stars and mass loss may have implications for the angular momentum evolution 
on the HB.

\begin{figure}
  \includegraphics[height=.3\textheight]{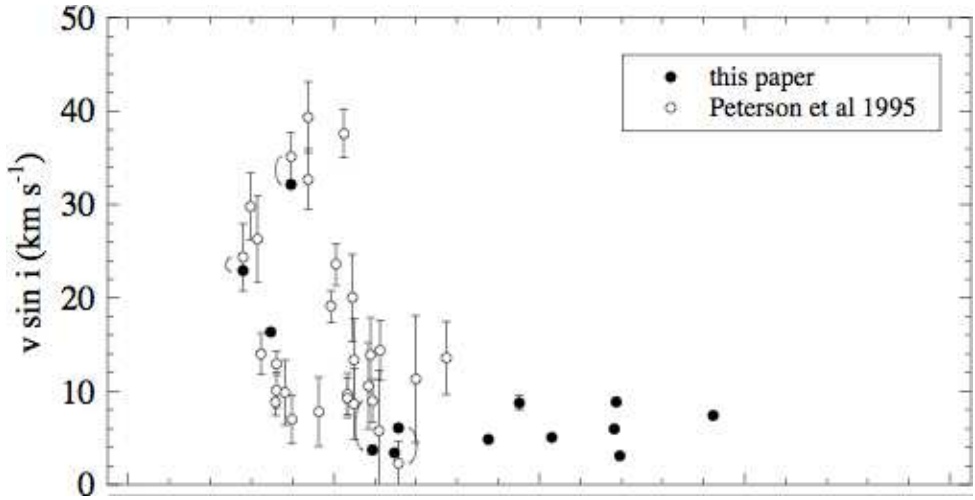}
  \caption{Rotational velocities of HB stars as a function of effective temperature. Note that all BHB stars hotter than 
   $\sim$ 10~000 K are rotating slowly. Figure adapted from Behr et al. (2000).}
  \label{behrs}
\end{figure}

Figure~\ref{behrs} shows a compilation of rotational velocities of HB stars in globular clusters as a function of 
effective temperature (Behr et al. 2000). 
For stars cooler than $\sim$ 10 000 K, there is a spread of slow and rapid ($v$ sin$i$ $= 40$ km/s) 
rotators, whilst for temperatures hotter than $\sim$ 10 000 K, {\it all} HB stars rotate slowly 
(see also Brown \& Salaris, these proceedings). 
The temperature of this transition coincides with the onset of radiative diffusion where the Fe abundance 
of initially metal-poor BHB stars increases by a factor of 100. According to Eq.~\ref{eq_formula2}, the mass-loss rate may also 
increase by up to 2 orders of magnitude as a result.
Vink \& Cassisi (2002) argued that this large increase in mass-loss provides a natural explanation for 
the low rotational velocities of HB stars hotter than $\sim$ 10 000 K due to angular momentum loss. 

\section{Mass loss as a function of metallicity}

\begin{figure}
  \includegraphics[height=.3\textheight]{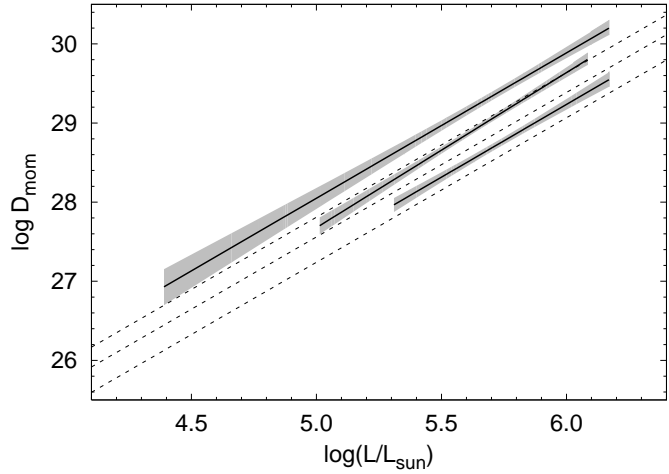}
  \caption{This figure represents a confrontation between O star mass-loss predictions (dashed line) and 
recent empirical mass-loss rates in the form of the so-called wind momentum-luminosity relationship for Galactic (top), 
LMC (middle) and SMC (bottom) O stars. The figure is from Mokiem et al. (2007).}
  \label{wlt_comp}
\end{figure}

Metallicity, $Z$, is a key parameter in the physics of stars, galaxies, and the cosmos, via 
metallicity-dependent stellar winds. Figure~\ref{wlt_comp} compares 
the predictions of O star mass-loss (Vink et al. 2001; dashed line) with recent 
empirical mass-loss estimates of Mokiem et al. (2007) in the form of the 
modified wind momentum-luminosity relationship, $D$, for Galactic, LMC, and SMC O stars in the $Z$ range solar to 1/5 
solar. The modified wind momentum is constructed from the multiplication of the rate of mass loss and the terminal wind velocity modified 
by the stellar radius (see Kudritzki \& Puls 2000).

It can be noted that for all three galaxies (Milky Way, LMC, SMC), the empirical rates are larger
than predicted by theory. As the empirical rates are most likely affected by wind clumping, the empirical
rates are likely maximal. If we assume a modest clumping factor with a corresponding reduction in the empirical $\dot{M}$ 
by a factor 2-3, the empirical rates show good agreement with the wind models. The wind clumping factor however
remains an unsolved problem in stellar astrophysics and if the true wind clumping is larger than assumed, with 
empirical $\dot{M}$ overestimates of $\sim$ 10 as some studies (e.g. Fullerton et al. 2006) suggest, then
the current mass-loss predictions might also be too large. This is an important topic for further investigation.

\begin{figure}
  \includegraphics[height=.3\textheight]{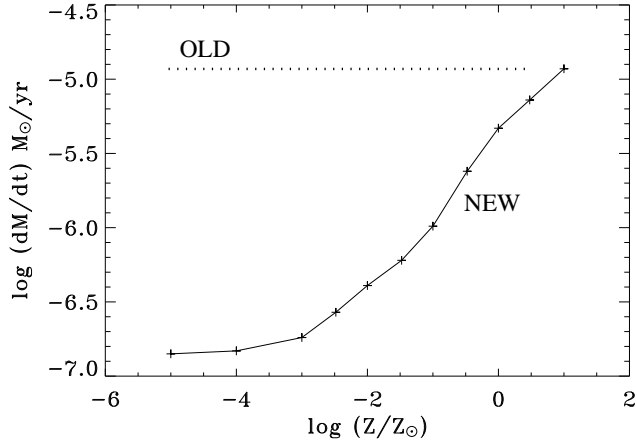}
        \caption{The first $Z$-dependent mass-loss predictions for Wolf-Rayet stars (Vink \& de Koter 2005).
Despite the overwhelming presence of carbon for all $Z$, the new WR mass-loss rate does {\it not} show
a $Z$-independent behaviour as assumed previously (see dotted line).
Instead, the new computations show that WR mass loss depends strongly on iron ($Z$) --- a {\it key}
result for predicting a high occurrence of long-duration GRBs at low metallicity.}
  \label{mdotz2}
\end{figure}

Massive stars lose mass at the most extreme rates during the Wolf-Rayet (WR) phase. 
In this phase, the outer layers become chemically enriched, which may modify mass loss 
through winds. 
We have recently performed a study of mass loss versus $Z$ for late-type
WR stars using the Monte Carlo approach. The outcome is shown in Fig.~\ref{mdotz2}.
Despite the overwhelming presence of carbon at all $Z$, $\dot{M}$ does {\it not} show 
a $Z$-independent behaviour (as usually assumed previously), but WR mass loss depends strongly on iron (Fe).

Furthermore, although the \mdot\ versus $Z$ dependence is consistent with a power-law decline in the observable 
Universe down to log $Z/\zsun \sim -3$, it flattens off for extremely low $Z$ models.
The reason is that carbon, nitrogen, oxygen, hydrogen and helium 
take over the driving due to iron (Fe) which is dominant at high $Z$.

The strong $Z$-dependence of WR winds shown in Fig.~\ref{mdotz2} where the WR $\dot{M}$ drops by orders of magnitude, 
represents a key result for the high frequency of long-duration GRBs at low metallicity. 
The favoured progenitors of long-duration gamma-ray bursts (GRBs) are rapidly rotating 
Wolf-Rayet (WR) stars, however, most Galactic WR stars are slow rotators, as stellar winds are thought to remove 
angular momentum, which may pose a challenge to the collapsar model for GRBs.
Observational data however indicate that GRBs occur predominately in low metallicity ($Z$) galaxies (e.g. Stanek et al. 2006), which may 
resolve the problem: lower $Z$ leads to less mass loss, which may inhibit  
angular momentum removal, allowing WR stars to remain rotating rapidly until collapse (Yoon \& Langer 2005, Woosley \& Heger 2006).

As a test of this scenario, 
Vink (2007) performed a linear spectropolarimetry survey of WR stars in the low $Z$ 
environment of the LMC and found an incidence of line polarisation effects in LMC WR stars as low as that of the Galactic sample of 
Harries et al. (1998). This suggests that the threshold metallicity where significant 
differences in WR rotational properties occur is below that of the LMC (at $Z$ $\sim$ 0.5 $\zsun$), possibly constraining GRB 
progenitor channels to this upper metallicity.

%\begin{theacknowledgments}
%I would like to thank my collaborators and friends, and in particular Rubina Kotak and Alex de Koter, for their support. 
%\end{theacknowledgments}



\bibliographystyle{mn2e}      % Modified 

\bibliography{vink_cam}

\IfFileExists{\jobname.bbl}{}
 {\typeout{}
  \typeout{******************************************}
  \typeout{** Please run "bibtex \jobname" to optain}
  \typeout{** the bibliography and then re-run LaTeX}
  \typeout{** twice to fix the references!}
  \typeout{******************************************}
  \typeout{}
 }

\end{document}